\documentclass[prd,twocolumn,twoside,preprintnumbers,superscriptaddress,nofootinbib]{revtex4}

\usepackage{amsmath}
\usepackage{graphicx}
\usepackage{dcolumn}
\usepackage[hyperfootnotes=false]{hyperref}
\usepackage{xspace}
\usepackage{color}

\newcommand{\fig}[1]{Fig.~\ref{#1}}

\newcommand{\muu}{m_u}
\newcommand{\md}{m_d}

\newcommand{\MeV}{\,\text{MeV}}
\newcommand{\GeV}{\,\text{GeV}}
\newcommand{\Br}{\text{Br}}

\begin{document}
\preprint{\vbox{\hbox{CERN-PH-TH-2014-068}}}

\title{Improved predictions for $\boldsymbol{\mu\to e}$ conversion in nuclei\\ and Higgs-induced lepton flavor violation}

\author{Andreas Crivellin}
\affiliation{CERN Theory Division, CH--1211 Geneva 23, Switzerland}%
\author{Martin Hoferichter}
\author{Massimiliano Procura}
\affiliation{Albert Einstein Center for Fundamental Physics,
Institute for Theoretical Physics,\\ University of Bern, Sidlerstrasse 5, CH--3012 Bern, Switzerland}

\begin{abstract}

Compared to $\mu \to e \gamma$ and $\mu \to e e e$, the process $\mu \to e$ conversion in nuclei receives enhanced contributions from Higgs-induced lepton flavor violation. Upcoming $\mu \to e$ conversion experiments
with drastically increased sensitivity will be able to put extremely stringent bounds on Higgs-mediated $\mu \to e$ transitions.
We point out that the theoretical uncertainties associated with these Higgs effects, encoded in the couplings of quark scalar operators to the nucleon, can be accurately assessed using our recently developed approach based on $SU(2)$ Chiral Perturbation Theory that cleanly separates two- and three-flavor observables.
We emphasize that with input from lattice QCD for the coupling to strangeness $f_s^N$, hadronic uncertainties are appreciably reduced compared to the traditional approach where $f_s^N$ is determined from the pion--nucleon $\sigma$-term by means of an $SU(3)$ relation. We illustrate this point by considering Higgs-mediated lepton flavor violation in the Standard Model supplemented with higher-dimensional operators, the two-Higgs-doublet model with generic Yukawa couplings, and the Minimal Supersymmetric Standard Model.
Furthermore, we compare bounds from present and future $\mu \to e$ conversion and $\mu \to e \gamma$ experiments.

\end{abstract}
%

\maketitle

\section{Introduction}
\label{intro}

Flavor-changing neutral current processes are strongly suppressed in the Standard Model (SM) and therefore sensitive even to small new physics (NP) contributions. Lepton flavor violation (LFV) is an especially promising probe of NP since in the SM with massive neutrinos all flavor-violating effects in the charged lepton sector are proportional to tiny neutrino masses.\footnote{For a review we refer to~\cite{Raidal:2008jk}.} For instance, the decay rates of heavy charged leptons into lighter ones are suppressed at least by $m_\nu^4/m_W^4$, where $m_\nu$ ($m_W$) is the neutrino ($W$-boson) mass. This leads to branching ratios of the order of $10^{-50}$, which are thus by far too small to be measurable in any foreseeable experiment. Therefore, any evidence of charged LFV would be a clear signal of physics beyond the SM. 

Different LFV observables are sensitive to distinct combinations of higher-dimensional operators and can therefore distinguish between various models of NP. Processes involving $\mu\to e$ transitions are especially powerful in discriminating NP scenarios since three different channels with excellent experimental sensitivity are available:\footnote{For a review on muon physics and BSM searches, see e.g.~\cite{Kuno:1999jp}.} $\mu \to e \gamma$, $\mu \to e e e$, and $\mu \to e$ conversion in nuclei. They provide at present the strongest constraints on LFV. The current limits from the first two processes are ${\Br}\,[\mu \to e \gamma] < 5.7 \times 10^{-13}$~\cite{Adam:2013mnn} and ${\Br}\,[\mu \to e e e] < 1.0 \times 10^{-12}$~\cite{Bellgardt:1987du}.
Moreover, $\mu \to e$ conversion in nuclei --- despite hadronic uncertainties --- is expected to be the best channel to experimentally test charged LFV due to the distinctive feature that the energy of the conversion electron lies well above the energy of the particles from ordinary muon decay~\cite{Mihara:2013zna}, contrary to $\mu \to e \gamma$ and $\mu \to e e e$. The current limit on the conversion rate normalized to the muon capture rate
\begin{equation} \label{rate_ex}
{\Br}_{\mu \to e} \equiv \frac{\Gamma \,(\mu^- N \to e^- N)}{\Gamma \,(\mu^- N \to {\text{nuclear capture}})}
\end{equation}
 is set by the SINDRUM~II experiment at PSI~\cite{Bertl:2006up} (with gold target nuclei)
\begin{equation}
{\Br}_{\mu \to e}^{\rm Au} \leq 7 \times 10^{-13}
\end{equation}
at $90 \%$ confidence level. The DeeMe experiment aims at an accuracy of $10^{-14}$~\cite{Aoki:2012zza}, while in the future Mu2e at FNAL and COMET at J-PARC~\cite{Carey:2008zz,Kutschke:2011ux,Cui:2009zz} are expected to improve the bound on the conversion rate by four orders of magnitude compared to SINDRUM~II. Since muon conversion is a coherent process, ${\Br}_{\mu \to e}$ in~\eqref{rate_ex} is proportional to the atomic number $Z$ in the case of isospin-conserving NP. Furthermore, dipole, vector, and scalar operators contributing to $\mu \to e$ conversion (see~\eqref{Leff} below) exhibit different sensitivities to $Z$. Therefore, a combined phenomenological analysis based on different target nuclei can help discriminate among different isospin-violating NP models. 

$\mu \to e$ conversion is especially important in the context of Higgs-mediated LFV because $\mu \to e \gamma$ and $\mu \to e e e$ are suppressed by small Yukawa couplings in this scenario. Higgs-induced LFV occurs in many NP models, such as the two-Higgs-doublet model (2HDM) or the Minimal Supersymmetric Standard Model (MSSM). Furthermore, the Higgs is intrinsically related to flavor physics, motivating a study of the corresponding higher-dimensional operator that generates flavor-changing SM Higgs couplings~\cite{Goudelis:2011un,Blankenburg:2012ex,Harnik:2012pb}. 

The effects of Higgs exchange are encoded in scalar couplings to the nucleon~\cite{Cirigliano:2009bz,Gonzalez:2013rea}. While the precision of the forthcoming experiments calls for a careful analysis of the ensuing hadronic uncertainties,\footnote{Further hadronic input is required for the overlap integrals between the electron and muon wave functions with the nucleon densities, which have been calculated in~\cite{Kitano:2002mt}.}
 these couplings are often extracted~\cite{Ellis:2000ds,Corsetti:2000yq} using an empirical formula based on soft flavor $SU(3)$ symmetry breaking~\cite{Cheng:1988im}. 
 In particular, in this approach the scalar couplings involving $u$- and $d$-quarks are reconstructed from three-flavor quantities, thus misrepresenting the actual uncertainties due to unnecessary $SU(3)$ assumptions. 
 In this paper we use the nucleon scalar couplings to $u$- and $d$-quark as determined in~\cite{Crivellin:2013ipa}, relating two-flavor dependent quantities to phenomenology in a rigorous, model-independent way based on $SU(2)$ Chiral Perturbation Theory (ChPT), 
 and quantify the effects on $\mu \to e$ conversion in nuclei.

In Sect.~\ref{sec:mueconv} we provide all formulae relevant for an accurate phenomenological analysis, focusing on the role of hadronic uncertainties. In Sect.~\ref{sec:Higgs-mediatedFV} we work out the corresponding constraints on the SM dimension-$6$ effective operator generating Higgs-mediated LFV, and discuss the specific cases of the MSSM and the 2HDM.


\section{The $\boldsymbol{\mu\to e}$ conversion rate}
\label{sec:mueconv}


Our starting point is the effective Lagrangian below the electroweak-symmetry-breaking scale containing all operators that contribute to coherent $\mu\to e$ conversion in nuclei up to dimension $8$, with at most four SM fields (see e.g.~\cite{Cirigliano:2009bz})
\begin{align} \label{Leff}
{\cal L}_{\rm{eff}}=\;& {{m_\mu } \Big( {C_T^R\,\bar e{\sigma ^{\rho \nu }}{P_L}\mu  + C_T^L\,\bar e{\sigma ^{\rho \nu }}{P_R}\mu } \Big){F_{\rho \nu }}} \nonumber \\
& + \Big( {C_{qq}^{V L}\,\bar e{\gamma ^\nu }{P_L}\mu  + C_{qq}^{V R}\,\bar e{\gamma ^\nu}{P_R}\mu } \Big)\,\bar q{\gamma _\nu }q \nonumber \\
& +{\Big( {C_{qq}^{S R}\,\bar e{P_L}\mu  + C_{qq}^{S L}\,\bar e{P_R}\mu } \Big )\,m_\mu {m_q}\bar qq} \nonumber \\
& + m_\mu\, {\alpha _s}\Big( {C_{gg}^R\,\bar e{P_L}\mu  + C_{gg}^L\,\bar e{P_R}\mu } \Big)G_{\rho \nu }G_{}^{\rho \nu }\,.
\end{align}
Here $F_{\rho \nu}$ and $G_{\rho \nu}$ are the electromagnetic and the gluon field strength tensors, respectively, $P_{R,L}=(1\pm\gamma_5)/2$ denotes the chiral projectors, and $q=u,d,s,c,b,t$ any quark flavor. The dimensionful Wilson coefficients $C$ are suppressed by increasing powers of the NP scale $\Lambda$. We defined them to be renormalization-group invariant (at one loop under QCD).

At the scale where the nucleon matrix elements are evaluated only the light quarks ($u$, $d$, and $s$) are dynamical degrees of freedom, while the contribution from heavy quarks is absorbed into a redefinition of the Wilson coefficient of the gluon operator, by calculating threshold corrections and using the trace anomaly~\cite{Shifman},
\begin{equation}
C_{gg}^L \to C_{gg}^L - \frac{1}{{12\pi }}\sum\limits_{Q = c,b,t} {C_{QQ}^{S L}}\,,
\end{equation}
and the same for $L$ replaced by $R$. The conversion rate is given by
\begin{align} \label{convrate}
 \Gamma_{\mu \to e} =& \frac{m_\mu ^5}{4}\,\Big| C_T^L \,D
 + 4\Big[ m_\mu \,m_p \Big( \tilde C_p^{SL} - 12\pi\, \tilde C_p^{ggL} \Big)\,{S^p}
 \nonumber \\ 
&+ \tilde C_p^{VL}\,{V^p} + p \to n \Big] \Big|^2 + L \to R\, ,
\end{align}
with $p$ and $n$ denoting proton and neutron, respectively. The dimensionless coefficients $D$, $S^N$, and $V^N$ are related to the overlap integrals of the initial-state muon $1 S$ wave function and the final-state electron wave function with the target nucleus. For the numerical estimate of these input parameters we use the outcome of~\cite{Kitano:2002mt},\footnote{The explicit numbers are: $D=0.189$, $S^p=0.0614$, $V^p=0.0974$, $S^n=0.0918$, $V^n=0.146$ for gold, and $D=0.0362$, $S^p=0.0155$, $V^p=0.0161$, $S^n=0.0167$, $V^n=0.0173$ for aluminum targets.}
which followed the approach in~\cite{Czarnecki:1998iz}. In addition, we defined
\begin{align}
\tilde C_p^{VR} &= \sum\limits_{q = u,d} {C_{qq}^{VR}} f_{{V_q}}^p\,,\notag\\
\tilde C_p^{SR} &= \sum\limits_{q = u,d,s} {C_{qq}^{SR}} f_q^p\,,\notag\\
\tilde C_p^{ggL} &= C_{gg}^L\,f_Q^p\,,
\end{align}
which account for the quark content of the proton via the vector (scalar) couplings $f_{{V_q}}^N$ ($f_q^N$). The analogous equations for left-handed operators are obtained by replacing $R$ with $L$, and those for the neutron by substituting $p$ with $n$.  

In the numerical analysis we will use the values for the $u$- and $d$-quark scalar couplings $f_q^N$ derived in~\cite{Crivellin:2013ipa} in the framework of two-flavor ChPT. This approach avoids relying on three-flavor input for these two-flavor quantities, contrary to the procedure often applied in the literature~\cite{Ellis:2000ds,Corsetti:2000yq,Cirigliano:2009bz} where the strangeness content of the nucleon $y$ is combined with another parameter to reconstruct the two-flavor couplings. The latter quantity, which measures the amount of isospin breaking, is usually taken from leading-order fits to the baryon spectrum~\cite{Cheng:1988im}, which further hinders a proper error estimate due to unknown higher-order corrections. 
Our approach allows us to directly connect $f_u^N$ and $f_d^N$ to the pion--nucleon $\sigma$-term $\sigma_{\pi N}$ and consistently account for isospin-breaking effects, which were overestimated by a factor of $2$ in the traditional approach~\cite{Crivellin:2013ipa}. Summarizing the findings of~\cite{Crivellin:2013ipa}, the $u$- and $d$-couplings are given by
\begin{align}
\label{fq_res}
f_u^N&=\frac{\sigma_{\pi N}(1-\xi)}{2m_N}+\Delta f_u^N\,,\quad\!\!
f_d^N=\frac{\sigma_{\pi N}(1+\xi)}{2m_N}+\Delta f_d^N\,,\notag\\
  \Delta f_u^p
 &=(1.0\pm 0.2)\cdot 10^{-3}\,,\quad 
   \Delta f_u^n
 =(-1.0\pm 0.2)\cdot 10^{-3}\,,\notag\\
   \Delta f_d^p
 &=(-2.1\pm 0.4)\cdot 10^{-3}\,,\quad
   \Delta f_d^n
 =(2.0\pm 0.4)\cdot 10^{-3}\,,\notag\\
\xi&=\frac{\md-\muu}{\md+\muu}=0.36\pm 0.04\,, 
\end{align}
with $\muu/\md=0.47\pm 0.04$ taken from~\cite{FLAG}. We refrain from quoting a range for the $\sigma$-term, but rather
express our results as a function of $\sigma_{\pi N}$.\footnote{For a compilation of lattice results for $\sigma_{\pi N}$ we refer to~\cite{Young,Kronfeld:2012uk,WalkerLoud,Belanger:2013oya}, for phenomenological determinations to~\cite{Gasser:1990ce,Pavan:2001wz,Alarcon:2011zs}. Its extraction from $\pi N$ scattering can be improved using the precision measurements of the scattering lengths in pionic atoms~\cite{Gotta:2008zza,Strauch:2010vu,piD,piDlong}, as well as constraints from analyticity and crossing symmetry~\cite{RS,RSSFF}.}  

The scalar coupling to the $s$-quark $f_s^N$ and the strangeness content of the nucleon $y$ are related via
\begin{equation}
\label{yfs}
y= \frac{m_Nf_s^N}{\sigma_{\pi N}} \frac{2\hat m}{m_s}\,,
\end{equation}
where $\hat m=(\muu+\md)/2$ and $m_s/\hat m=(27.4\pm 0.4)$~\cite{FLAG}. Traditionally, $y$ has been derived from the $SU(3)$ relation $y=1-\sigma_0/\sigma_{\pi N}$, with $\sigma_0=(36\pm 7)\MeV$~\cite{Borasoy}. However, this approach is very sensitive to the precise value of $\sigma_{\pi N}$, and depending on the $\sigma$-term input has led to large values $f_s^N\approx 0.25$ that in view of recent lattice calculations, with $m_s$ close to the physical point, appear increasingly unlikely.  
Therefore, in this paper we use the lattice average from~\cite{WalkerLoud}
\begin{equation}
\label{fs_lattice}
f_s^N=0.043\pm 0.011\,.
\end{equation}
A narrower range for $y$ motivated by lattice calculations was already used in~\cite{Cirigliano:2009bz}. Since the translation to $f_s^N$ according to~\eqref{yfs} again involves $\sigma_{\pi N}$, we prefer to use $f_s^N$ directly. 

At leading order in $\alpha_s$ the scalar couplings for the heavy quarks are the same, and equal to~\cite{Shifman}\footnote{For a discussion of $f_Q^N$ at higher orders in $\alpha_s$ see~\cite{Kryjevski,Vecchi}.}
\begin{align}
 f_Q^N&=\frac{2}{27}\big(1-f_u^N-f_d^N-f_s^N\big).
\end{align}

\begin{figure}
\includegraphics[height=45ex]{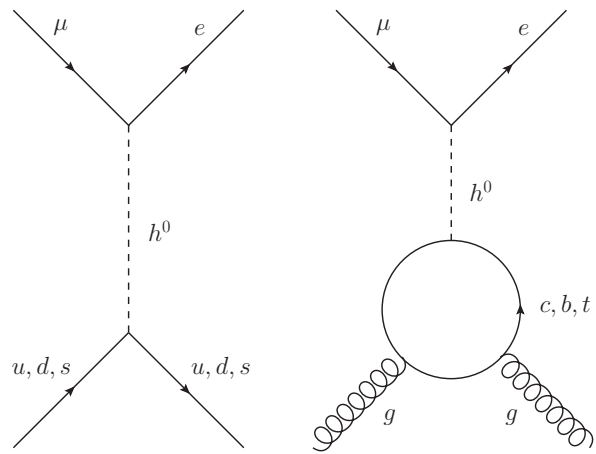}%
\caption{$\mu\to e$ conversion in the case of Higgs-induced LFV. The left diagram shows the tree-level contribution involving light quarks, while the right diagram depicts the loop-induced effect of heavy quarks to the gluon operator.}
\label{diagrams}
\end{figure}

In the next section we focus on Higgs-induced LFV, which naturally generates scalar interactions, see Fig.~\ref{diagrams}. We explore the constraints set by experimental limits both for dimension-$6$ effective operators and in the framework of the 2HDM, and study the impact of different assumptions for the scalar couplings on the estimate of hadronic uncertainties associated with $\mu \to e$ conversion. 

\section{Higgs-mediated flavor violation}
\label{sec:Higgs-mediatedFV}

LFV processes have been studied in great detail in
many extensions of the SM. For example, in the MSSM
non-vanishing rates for LFV processes are generated by flavor
non-diagonal SUSY-breaking terms~\cite{Borzumati:1986qx,
  Brignole:2004ah, Paradisi:2005fk,Paradisi:2006jp,Altmannshofer:2009ne,
  Girrbach:2009uy}. Extending the MSSM with right-handed
neutrinos by the seesaw mechanism~\cite{Minkowski:1977sc} gives rise
to LFV~\cite{Ilakovac:1994kj,Hisano:1995cp, Hisano:2001qz,
  Babu:2002et, Masiero:2004js, Dedes:2007ef, Antusch:2006vw,
  Ilakovac:2012sh,Ilakovac:2013wfa}, as well as allowing for $R$-parity
violation~\cite{deGouvea:2000cf, Abada:2001zh, Dedes:2006ni}. The Littlest Higgs Model with $T$-parity~\cite{Blanke:2007db}, 2HDMs with generic flavor
structures~\cite{Kanemura:2004cn, Kanemura:2005hr, Paradisi:2005tk,Crivellin:2013wna}, and models with an extended fermion sector~\cite{Buras:2011ph} have sources of LFV as well.  In
order to make NP scenarios consistent with the
non-observation of LFV processes in nature, the assumption of Minimal
Flavor Violation~\cite{MFV} has been extended to the lepton sector, see e.g.~\cite{Cirigliano:2005ck, Nikolidakis:2007fc}. 
LFV decays have been studied in a model-independent way in~\cite{Raidal:1997hq,Cirigliano:2009bz,Dassinger:2007ru,Crivellin:2013hpa,Celis:2014asa}.


In this section we consider three cases where LFV is induced via Higgs exchange: the SM extended with a dimension-$6$ operator, the 2HDM, and the MSSM. In general, for the exchange of any number of neutral scalar particles $H_k^0$ (i.e.\ SM Higgs and possible new scalars in various BSM scenarios), the quark scalar Wilson coefficients in~\eqref{Leff} receive the following tree-level contributions
\begin{align}
C_{qq}^{S R} &= \frac{{{1}}}{{{m_q}{m_\mu }}}\sum\limits_k {\frac{{{\mathop{\rm Re}\nolimits} \left[ {\Gamma _{qq}^{H_k^0\;LR}} \right]\Gamma _{\mu e}^{H_k^0\;LR*}}}{{m_{H_k^0}^2}}}\,, \notag\\
C_{qq}^{S L} &= \frac{{{1}}}{{{m_q}{m_\mu }}}\sum\limits_k {\frac{{{\mathop{\rm Re}\nolimits} \left[ {\Gamma _{qq}^{H_k^0\;LR}} \right]\Gamma _{e\mu }^{H_k^0\;LR}}}{{m_{H_k^0}^2}}} \,.
\end{align}
Here $k$ runs over the number of neutral scalars in the theory and the scalar--fermion--fermion terms in the NP Lagrangian are $\Gamma_{q_i q_j}^{H_k^0\, LR}\bar q_i P_R q_j H^0_k +\text{h.c.}$ and $\Gamma_{\ell_i \ell_j}^{H_k^0\, L R}\bar \ell_i P_R \ell_j H^0_k+\text{h.c.}$, where $i$ and $j$ label different flavors, i.e.\ $\ell_1=e$, $\ell_2=\mu$, $\ell_3=\tau$, etc. Since masses have the same anomalous dimension under QCD as the couplings $\Gamma_{i j}^{H_k^0}$ (assuming that the scalar particles are not strongly-interacting), both quantities should be taken at the same scale in the numerical analysis. 

For the comparison of constraints from $\mu\to e$ conversion with $\mu\to e\gamma$
we also need
\begin{align}
\label{muegBR}
\Br\left[ {\mu  \to e\gamma } \right] = \frac{{m_\mu ^5}\,{\tau_\mu }}{{4\pi }}\left( {|C_T^L{|^2} + |C_T^R{|^2}} \right){\mkern 1mu}\,.
\end{align}
Here, $\tau_\mu $ is the muon's mean lifetime and, for a general number of neutral Higgs bosons,
\begin{align}
C_T^L &= \frac{e}{192\pi ^2}\sum\limits_k \frac{1}{m_{H_k^0}^2}\Bigg\{ \Gamma _{e\mu }^{H_k^0\;LR}\Big( \Gamma _{\mu \mu }^{H_k^0\;LR \star} + \Delta _{\text{2-loop}}^{\text{Barr--Zee}} \Big) 
\nonumber \\
&- \sum\limits_{\ell  = e,\mu ,\tau } \frac{m_\ell }{m_\mu }\Gamma _{e\ell }^{H_k^0\;LR}\Gamma _{\ell \mu }^{H_k^0\;LR}\bigg[ 9 + 6\log {\frac{m_\ell ^2}{m_{H_k^0}^2}} \bigg] \Bigg\} 
\label{CTL}
\end{align}
in the notation of~\cite{Crivellin:2013wna}. The analogous expression for $C_T^R$ is simply obtained by replacing $L$ with $R$.
$\Delta _{\text{2-loop}}^{{\text{Barr--Zee}}}$ denotes the contribution from Barr--Zee-type diagrams at two loops~\cite{Barr:1990vd}, which are known to be numerically very important due to an $m_t^2/m_\mu^2$ enhancement. In the context of SM
flavor-changing Higgs processes (where the muon Yukawa coupling is not
$\tan \beta$-enhanced), they even give the dominant effect in $\mu \to e
\gamma$~\cite{Chang:1993kw,Blankenburg:2012ex,Harnik:2012pb}.

\subsection{Flavor-changing SM Higgs couplings from higher-dimensional effective operators}

We first consider an extension of the SM with a dimension-$6$ operator giving rise to flavor-changing couplings of the SM Higgs. This arguably provides the most transparent way to illustrate how hadronic input for the scalar couplings to the nucleon affects the theory predictions. There is only one operator of dimension $6$ which is not stringently constrained from $\mu \to e \gamma$ or $\mu \to eee$~\cite{Crivellin:2013hpa} and can induce sizable LFV SM Higgs couplings, namely
\begin{align}
\mathcal{O}_{e\phi}^{ij} = \left( {{\phi ^\dag }\phi } \right)\left( {{{\bar \ell }_i}{e_j}\phi } \right)\,.
\end{align}
Here $\phi$ refers to the SM Higgs doublet, $\ell_i$ to a lepton doublet of flavor $i$, and $e_j$ to a lepton singlet of flavor $j$.  
The Higgs--lepton--lepton coupling (in the physical basis with diagonal mass matrices) in this case is given by
\begin{equation} \label{Gamma_H}
\Gamma _{\ell _f \ell _i }^{h^0 }  =  - \frac{{m_{\ell _i } }}{v} +
\frac{1}{{\sqrt 2 }}\frac{{v^2 }}{{\Lambda ^2 }}\tilde C_{e\phi}^{fi}\,.
\end{equation}
Here $v=246$ GeV and the mass matrix,
\begin{equation}
m_{fi}^\ell   = \frac{v}{{\sqrt 2 }}\left( {Y_{fi}^\ell   -
\frac{1}{2}\frac{{v^2 }}{{\Lambda ^2 }}C_{e\phi}^{fi} } \right)\,,
\end{equation}
has been diagonalized via the transformation 
\begin{equation}
U_{fj}^{\ell L\dag } m_{jj'}^\ell  U_{j'i}^{\ell R}  = m_{\ell _i }
\delta _{fi}\,,
\end{equation}
which defines the Wilson coefficients in~\eqref{Gamma_H}
\begin{equation}
\tilde C_{e\phi}^{fi}  = \left( {U_\ell ^{L\dag } C_{e\phi} U_\ell ^R } \right)_{fi}\,.
\end{equation}
The effect of this operator on LF observables and LFV Higgs decays has been studied in~\cite{Goudelis:2011un,Blankenburg:2012ex,Harnik:2012pb}.

We have to deal with two free parameters $C_{e\phi}^{12}$ and $C_{e\phi}^{21}$ (corresponding to $\Gamma_{\mu e}$ and $\Gamma_{e\mu}$), which allows us to point out in the most transparent way the differences between our approach for the nucleon scalar couplings and the approach based on~\cite{Cheng:1988im,Ellis:2000ds} and commonly used in the literature, see~\cite{Cirigliano:2009bz,Harnik:2012pb}.

\begin{figure}
\includegraphics[height=42ex]{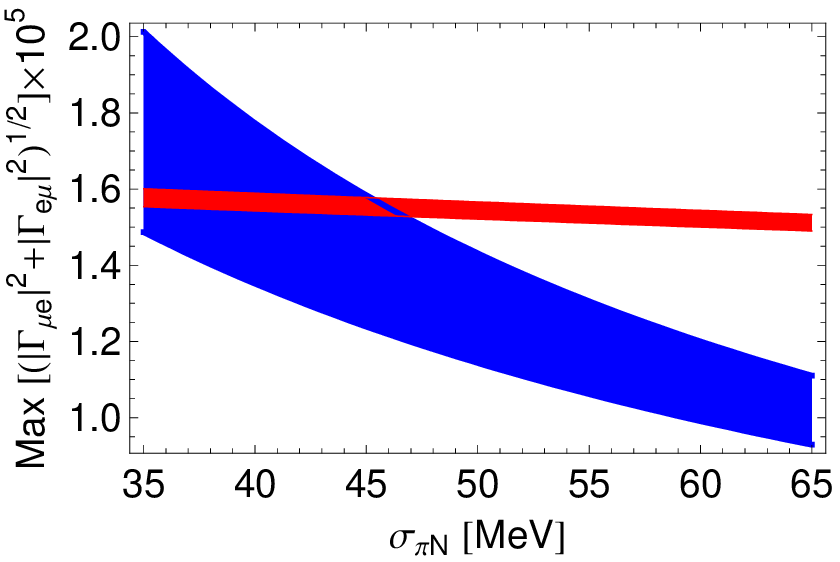}%
\hfill%
\includegraphics[height=40ex]{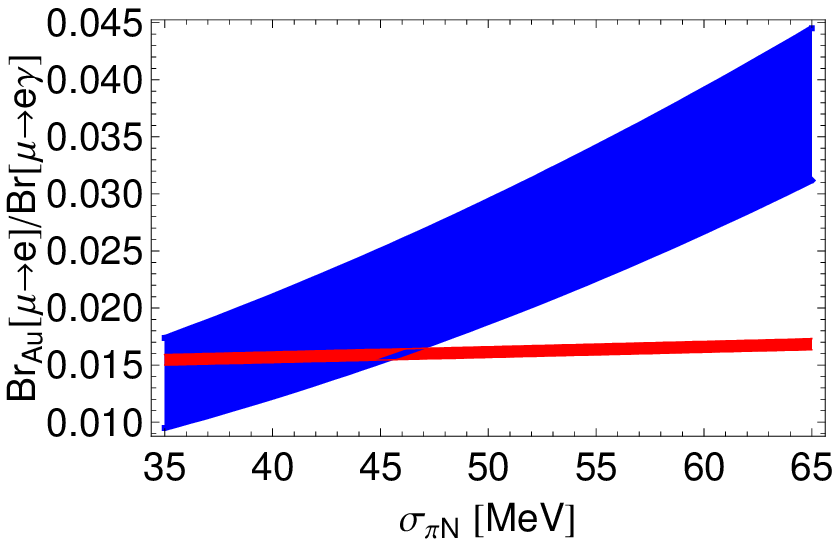}%
\caption{Upper panel: bounds on LFV SM Higgs couplings from $\mu \to e$ conversion in gold nuclei, as a function of the pion--nucleon $\sigma$-term. The blue band corresponds to the traditional way of assessing hadronic uncertainties, the red band to our approach where $f_s^N$ is determined by lattice QCD. Lower panel: predicted ratio of $\mu \to e$ conversion in gold nuclei and $\mu \to e \gamma$ for the two different approaches.}
\label{SMplots}
\end{figure}

The upper panel in \fig{SMplots} shows the bounds for $(|\Gamma_{\mu e}|^2+ |\Gamma_{e \mu}|^2)^{1/2}$ from $\mu\to e$ conversion in aluminum as a function of $\sigma_{\pi N}$.\footnote{We do not include the Barr--Zee-type diagrams for the vector operators in~\eqref{convrate}.
The corresponding offshell-photon contribution has not been calculated in the literature so far, but we expect 
it to be negligible for the following reasons:
first, it is suppressed by an additional factor $e$ compared to the magnetic-moment operator, and the wave-function overlap is about a factor of $2$ smaller. Most importantly, since the vector operator is chirality-conserving, the LFV coupling always has to be paired with an insertion of the muon mass, in contrast to the chirality-changing magnetic-moment operator. Therefore, the two-loop vector operator can be enhanced at most by $m_t/m_\mu$ compared to the one-loop magnetic-moment contribution.} 
For the numerical analysis the overlap integrals are taken from Table~I in~\cite{Kitano:2002mt} and the capture rate on gold atoms is the one determined in~\cite{Suzuki:1987jf}. We set $m_{h^0}=125$ GeV, $ \Delta _{\text{2-loop}}^{{\text{Barr--Zee}}} = -1.32\,m_\tau/m_\mu=-22.34$ using the results in~\cite{Harnik:2012pb}. The blue band in \fig{SMplots} corresponds to the traditional approach where the strangeness content $y$ is derived from the $SU(3)$ relation $y=1-\sigma_0/\sigma_{\pi N}$, and uncertainties estimated as explained in~\cite{Crivellin:2013ipa}.\footnote{Our analytic results agree with~\cite{Harnik:2012pb}, but our numerical bound on $(|\Gamma_{\mu e}|^2+ |\Gamma_{e \mu}|^2)^{1/2}$ is stronger than the one quoted in Table~I therein.
We thank Jure Zupan for rechecking the numerics of~\cite{Harnik:2012pb} and confirming our result.}
The difference in slope is due to the artificial $\sigma_{\pi N}$-dependence in this approach. 
Since only the sum of $f_u^N+f_d^N$ enters, the bulk of the isospin-violating corrections from~\eqref{fq_res} actually drops out. Moreover, 
the red band is remarkably stable against variations of $\sigma_{\pi N}$ and despite the large discrepancy of $f_s^N$ between both approaches the resulting effect on the bound is relatively moderate. The reason for this behavior can be understood from the fact that the relevant combination of scalar couplings
\begin{align}
&f_u^N+f_d^N+f_s^N+3\frac{2}{27}\big(1-f_u^N-f_d^N-f_s^N\big)\notag\\ 
&=\frac{2}{9}+\frac{7}{9}\big(f_u^N+f_d^N+f_s^N\big) 
\end{align}
is actually dominated by the constant term, e.g.\ for $\sigma_{\pi N}=50\MeV$ and $f_s^N$ according to~\eqref{fs_lattice} the second term merely amounts to $1/3$ of the first. 

The impact of taking $y$ from lattice or from the $SU(3)$ relation was already studied in~\cite{Cirigliano:2009bz}, and the theoretical uncertainties were found not to constitute a limiting factor in the model-discriminating power of $\mu\to e$ conversion experiments. Our findings show that if $f_s^N$ is used as input instead of $y$, the results become remarkably insensitive to $\sigma_{\pi N}$.
Hence, with $f_s^N$ taken from lattice, the artificial dependence on $\sigma_{\pi N}$ in the traditional approach is avoided and
the hadronic uncertainties are appreciably reduced (with central values that differ significantly from the traditional ones for $\sigma_{\pi N} \gtrsim 50\MeV$). 
The same conclusions can be drawn from an analysis of the ratio of $\mu \to e$ conversion and $\mu \to e \gamma$ branching fractions, see Fig.~\ref{SMplots}.\footnote{Our conclusions hold true also for the case of an aluminum target, as chosen for both the Mu2e and COMET experiments.}

\begin{figure}
\includegraphics[height=42ex]{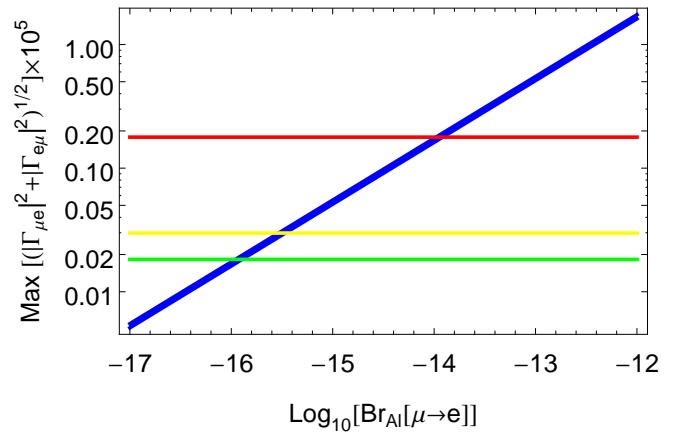}%
\caption{Bound on $(|\Gamma_{\mu e}|^2+ |\Gamma_{e \mu}|^2)^{1/2}$ as a function of the upper limit on $\mu\to e$ conversion in aluminum (blue band), compared to limits from $\mu\to e\gamma$ (horizontal lines). The red line refers to the present limit ${\Br}\,[\mu \to e \gamma] < 5.7 \times 10^{-13}$~\cite{Adam:2013mnn}, the yellow and green lines to future projections $1.6\times 10^{-14}$ and $6\times 10^{-15}$~\cite{Cheng:2013paa}.}
\label{muegVsconversion}
\end{figure}

The bound on $(|\Gamma_{\mu e}|^2+ |\Gamma_{e \mu}|^2)^{1/2}$ from $\mu\to e$ conversion is compared directly to the bound from $\mu\to e\gamma$ in Fig.~\ref{muegVsconversion}. In view of the upcoming experiments $\mu\to e$ conversion is likely to eventually provide the most stringent limits on Higgs-induced $\mu \to e$ transitions in the SM with dimension-$6$ operators.

\subsection{2HDM with generic Yukawa couplings}

In the 2HDM with generic Yukawa couplings (the 2HDM of type III) one has tree-level flavor-changing neutral Higgs couplings at tree level which can contribute to LFV processes, see e.g.~\cite{Kanemura:2004cn, Kanemura:2005hr, Paradisi:2005tk,Crivellin:2013wna}. In this case the general Higgs--fermion couplings of~\eqref{CTL} are given by
\begin{align}
{\Gamma_{u_f u_i }^{LR\, H_k^0} } &= x_u^k\left( \frac{m_{u_i }}{v_u/\sqrt{2}}
\delta_{fi} - \epsilon_{fi}^{u}\cot\beta \right) + x_d^{k\star}
\epsilon_{fi}^{u}\,, \notag\\
{\Gamma_{d_f d_i }^{LR\, H_k^0 } } &= x_d^k \left( \frac{m_{d_i
}}{v_d/\sqrt{2}} \delta_{fi} - \epsilon_{fi}^{d}\tan\beta \right) +
x_u^{k\star}\epsilon_{fi}^{ d} \,,\notag\\
{\Gamma_{\ell_f \ell_i }^{LR\,H_k^0} } &= x_d^k\left( \dfrac{m_{\ell_i }}{v_d/\sqrt{2}}
\delta_{fi} - \epsilon_{fi}^{\ell}\tan\beta \right) + x_u^{k\star} \epsilon_{fi}^{\ell}\,. 
\label{2HDM_couplings}
\end{align}
Here, $H^0_k=(H^0,h^0,A^0)_k$ refers to the heavy CP-even Higgs, the SM-like Higgs, and the CP-odd Higgs, respectively. The coefficients $x_q^{k}$ are given
by
\begin{align}
x_u^k  &=  \left(-\dfrac{1}{\sqrt{2}}\sin\alpha,\,-\dfrac{1}{\sqrt{2}}\cos\alpha,
\,\dfrac{i}{\sqrt{2}}\cos\beta\right)_k \,,\notag\\
x_d^k  &= \left(-\dfrac{1}{\sqrt{2}}\cos\alpha,\,\dfrac{1}{\sqrt{2}}\sin\alpha,
\,\dfrac{i}{\sqrt{2}}\sin\beta\right)_k \,.
 \end{align}
Assuming an MSSM-like Higgs potential the following relations among the parameters hold
\begin{align}
\label{Higgsmasses}
	\tan\beta&=\dfrac{v_u}{v_d}\,,\quad
	\tan 2\alpha=\tan 2\beta \; \dfrac{m^{2}_{A^{0}}+m_Z^2}{m^{2}_{A^{0}}-m_Z^2}\,,\notag\\
	m_{H^\pm}^2&=m^{2}_{A^{0}}+m_W^2 \,, \quad
	m_{H^0}^2=m^{2}_{A^{0}}+m_Z^2-m_{h^0}^2 \, .
\end{align}
The quantities $\epsilon^{q,\ell}_{i j}$ are the non-holomorphic Higgs--fermion couplings in the physical basis (see~\cite{Crivellin:2013wna} for details on the conventions). This means that $\epsilon^{d,\ell}_{i j}$ ($\epsilon^{u}_{i j}$) parametrize the coupling of down (up) quarks and leptons to the up- (down-)type Higgs doublet in the basis in which the fermion mass matrices are diagonal. In the limit $v \ll m_{A^0}$ all three non-SM Higgs masses in~\eqref{Higgsmasses} become equal $m_{A^0}\approx m_{H^\pm}\approx m_{H^0}\equiv m_H$.

\begin{figure}
\includegraphics[height=60ex]{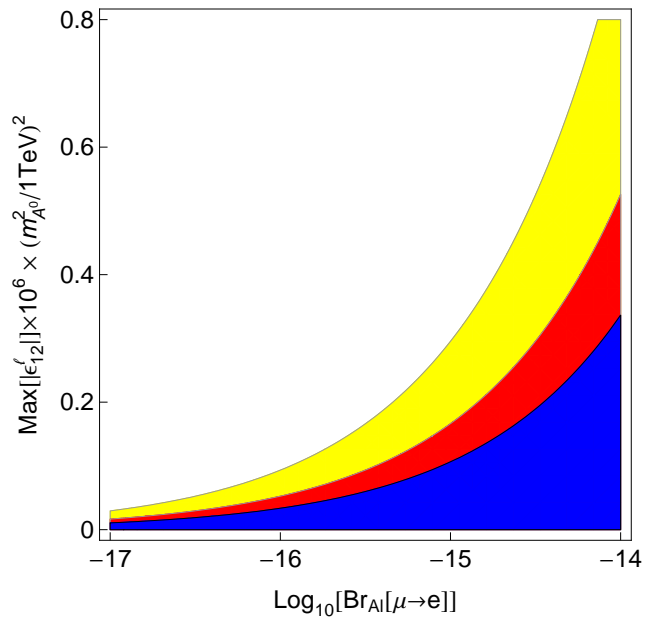}%
\caption{Allowed regions for $\epsilon^\ell_{12}\equiv\epsilon^\ell_{e\mu}$ as a function of the upper limit on $\mu\to e$ conversion in aluminum. The blue, red, and yellow regions correspond to $\tan \beta = 50, 40, 30$, respectively (the regions are superimposed with more stringent limits for larger $\tan\beta$). Note the simple quadratic scaling of the constraints on the heavy Higgs mass.}
\label{2HDMplot}
\end{figure}

We can now use $\mu\to e$ conversion to constrain the parameter space of the 2HDM~\cite{Paradisi:2006jp}. It is interesting to note that in the 2HDM the last term in~\eqref{CTL} drops out to a good approximation due to the cancellation between the heavy CP-even and the CP-odd Higgs contribution, while such a cancellation is absent in the tree-level contributions to $\mu\to e$ conversion. This further suppresses (in addition to the suppression by small Yukawa couplings) the $\mu \to e\gamma$ decay rate with respect to $\mu \to e$ conversion, and reinforces the expectation that these bounds will be more stringent than the ones from $\mu\to e \gamma$, as already observed for the case of the SM dimension-$6$ operator. As an illustration, in \fig{2HDMplot} we show the constraining power of $\mu \to e$ conversion as a function of the sensitivity within the reach of the future experiments with aluminum targets.

For simplicity we assumed an MSSM-like Higgs potential. This avoids CP violation and ensures at the same time unitarity and positivity of the Higgs potential. Of course there are also constraints from LHC searches and flavor observables on the 2HDM. For $\epsilon^\ell_{33}=0$ CMS currently excludes heavy Higges in the 2HDM with masses below $480\GeV$ for $\tan\beta=50$~\cite{Chatrchyan:2012vp}. This limit can be weakened if $\epsilon^\ell_{33}>0$. Concerning flavor constraints, $b\to s\gamma$ puts a lower limit on the charged Higgs mass of $380\GeV$~\cite{Hermann:2012fc} in the 2HDM of type II. This limit is to a very good approximation independent of $\tan\beta$ for $\tan\beta>1$ and can only be weakened by destructive interference originating from $\epsilon^u_{23}$. The tauonic $B$ decays $B\to \tau \nu$ and $B\to D^{(*)} \tau \nu$ recently showed some deviations from the SM predictions which cannot be accounted for in the 2HDM of type II~\cite{Lees:2012xj}. However, non-zero values of $\epsilon^u_{32,31}$ can bring bring experiment and theory predictions into agreement~\cite{Crivellin:2012ye}.

\subsection{The MSSM with heavy SUSY particles}

Many sources of LFV are present in the context of the MSSM (with or without seesaw mechanism)~\cite{Ilakovac:1994kj,Hisano:1995cp, Hisano:2001qz,
  Babu:2002et, Masiero:2004js,Paradisi:2006jp, Dedes:2007ef, Antusch:2006vw,Hisano:2010es,Ilakovac:2012sh,Ilakovac:2013wfa}. If the SUSY particles are heavier than the non-SM Higgses ($H^0,A^0$, and $H^\pm$), then the constraints on the couplings $\epsilon^{q,\ell}_{ij}$ obtained in the 2HDM (as discussed in the previous subsection) can be translated into bounds on SUSY-breaking parameters. Here we focus on this limit (i.e.\ $m_A^0\ll m_{\rm SUSY}$) and consider the region of moderate to large values of $\tan \beta$. 
  
The MSSM loop-contributions generating the parameters $\epsilon^{q,\ell}_{ij}$ of~\eqref{2HDM_couplings} have the important feature of being non-decoupling, i.e.\ they do not vanish in the limit of a large SUSY-breaking scale and depend only on ratios of SUSY parameters. Furthermore, loops generating $\epsilon^{q,\ell}_{ij}$ can be parametrically enhanced by $\tan\beta$~\cite{Banks:1987iu,Carena:1994bv,Carena:1999py,Isidori:2001fv,Hofer:2009xb} (and/or by $A^q/m_q$~\cite{Borzumati:1999sp,Crivellin:2010gw}, where $A^q$ is the trilinear SUSY-breaking term coupling squarks to the Higgs field). The Higgs exchange gives the dominant effect for large $\tan\beta$ if the additional heavy Higgses are lighter than the other SUSY particles.\footnote{If this hierarchy of masses is not realized, all loop effects contributing to $\epsilon^{q,\ell}_{ij}$ have to be taken into account (including boxes, $Z$-penguins, etc.)~\cite{Ilakovac:2013wfa}.} The complete one-loop expressions for $\epsilon^{q,\ell}_{ij}$ (in the decoupling limit) taking into account also the effects of the trilinear $A$-terms are given in~\cite{Crivellin:2011jt}.

Out of all possible Higgs--quark--quark couplings (that have to be inserted into the 2HDM expressions~\eqref{2HDM_couplings} to infer the MSSM contribution in the decoupling limit) only the down-quark couplings can get enhanced corrections compared to the tree-level expressions
\begin{equation}
	\epsilon^{d}_{ii}=\dfrac{\Sigma^{d\,LR}_{ii}}{\sqrt{2}\,v_u}\,.
\end{equation}
Here $\Sigma^{d\,LR}_{ii}$ is the part of the down-quark self-energy arising from the vacuum expectation value of the up-type Higgs doublet (see~\cite{Crivellin:2011jt} for details of the conventions). $\Sigma^{d\,LR}_{ii}$ can give a correction of about $50 \%$ to the corresponding quark mass for large values of $\tan\beta$~\cite{Banks:1987iu,Carena:1994bv,Carena:1999py,Isidori:2001fv,Hofer:2009xb}\footnote{For the NLO expressions we refer to~\cite{Bauer:2008bj,Bednyakov:2009wt,Noth:2010jy,Crivellin:2012zz}.} and is thus numerically very important.

At the loop level, both bilinear SUSY-breaking terms~\cite{Hamzaoui:1998nu} and $A$-terms~\cite{Crivellin:2010er} generate flavor-changing Higgs--lepton--lepton couplings. The leading term is proportional to $\tan^2\beta$ and involves only the flavor-changing element $\delta^{LL\,\ell}_{12}$ of the left-handed bilinear slepton terms, while $\delta^{RR\,\ell}_{12}$ enters only at a sub-leading level (suppressed by the ratio of gauge couplings $g_1^2/g_2^2$). $\delta^{RR\,\ell}_{ij}$ and $\delta^{LL\,\ell}_{ij}$ are the dimensionless off-diagonal elements of the slepton mass matrices normalized by the average squark mass. 

\begin{figure}
\includegraphics[height=60ex]{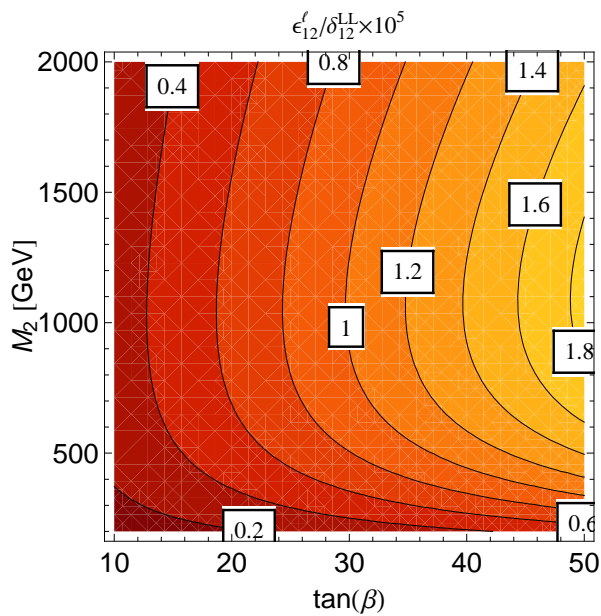}%
\caption{$\epsilon^{\ell}_{e \mu}$ normalized by the dimensionless parameter $\delta^{LL\,\ell}_{12}$ quantifying the amount of flavor violation in the slepton sector as a function of $\tan \beta$ and the Wino mass $M_2$.}
\label{MSSMplot}
\end{figure}

In the end, the dominant contribution to $\mu\to e$ conversion is due to down-quark operators and scales like $\tan^6\beta$ leading to stringent constraints~\cite{Hisano:1995nq,Hisano:2010es,Paradisi:2006jp} on the MSSM parameters. We show in Fig.~\ref{MSSMplot} the size of $\epsilon^{\ell}_{e \mu}\equiv\epsilon^{\ell}_{12}$ as a function of $\tan\beta$ and the mass $M_2$ of the Wino, which occurs together with a slepton in the loop that generates $\epsilon^{\ell}_{e \mu}$. Combining this with the constraints on the 2HDM parameter space (see Fig.~\ref{2HDMplot}), and taking into account the partially correlated effects in the Higgs--quark--quark couplings, one can obtain bounds on the MSSM parameter space.

\section{Conclusions}

$\mu \to e$ conversion is particularly sensitive to Higgs-mediated LFV because it is not suppressed by small Yukawa couplings as $\mu \to e \gamma $ and $\mu \to eee$ (and neither by a cancellation between the CP-odd and the CP-even Higgs contribution).
In this article we carefully investigated the impact of theoretical uncertainties induced by couplings of the nucleon to quark scalar operators in the context of $\mu \to e$ conversion in nuclei. 

The analysis relies on a clean separation of two- and three-flavor effects, with the $u$- and $d$-couplings expressed in terms of $\sigma_{\pi N}$ and isospin-breaking corrections derived in the rigorous framework of $SU(2)$ ChPT~\cite{Crivellin:2013ipa}.
This approach allows for a reliable assessment of uncertainties and thus a clarification of the role of hadronic uncertainties in $\mu\to e$ conversion. We find that isospin-breaking effects largely cancel, since only the sum of $u$- and $d$-quark enters, and that altogether the result is remarkably insensitive to variations of the $u$-, $d$-, and $s$-couplings, which can be traced back to a large constant term generated when integrating out the heavy quarks. We point out that taking the strangeness coupling $f_s^N$ from lattice calculations instead of determining $f_s^N$ from $\sigma_{\pi N}$ by means of an $SU(3)$ relation as often done in the literature, not only reduces hadronic uncertainties appreciably, but also removes a large artificial dependence on $\sigma_{\pi N}$.

 We applied our results for the hadronic quantities to the case where flavor-changing SM-Higgs couplings are induced by a dimension-$6$ operator. 
 Our bounds for the LFV couplings are stronger than previously thought. We further investigated the constraining power of future Mu2e and COMET experiments concerning flavor-changing parameters in the 2HDM, which can be translated into bounds on the MSSM parameter space. 
 In view of the forthcoming experiments, $\mu \to e$ conversion is likely to eventually provide the most stringent bounds on Higgs-mediated $\mu \to e$ transitions.

\section*{Acknowledgments}

We thank Roni Harnik, Gino Isidori, and Jure Zupan for helpful communication, Uli Nierste for useful discussions, and Jason Aebischer and Xavier Garcia i Tormo for comments on the manuscript. Support by the Swiss National Science Foundation and by the ``Innovations- und Kooperationsprojekt C-13" of the Schweizerische Universit{\"a}tskonferenz SUK/CRUS is gratefully acknowledged. A.C.\ is supported by a Marie Curie Intra-European Fellowship of the European Community's 7th Framework Programme under contract number (PIEF-GA-2012-326948). 

\bibliography{BIB_mueconv}

\end{document}